\documentclass[10pt, a4paper]{article}

\usepackage[utf8]{inputenc}
\usepackage[T1]{fontenc}
\usepackage{lmodern,textcomp}
\usepackage[english]{babel}
\usepackage{csquotes}

\usepackage[top=4cm, bottom=4cm, left=3.5cm, right=3.5cm]{geometry}

\usepackage{eurosym,xspace}
\usepackage[nottoc]{tocbibind}
\usepackage{graphicx,subfig,float}
\usepackage[multiple]{footmisc}
\usepackage{authblk}
\usepackage{fancyhdr}
\usepackage[backend=bibtex8, citestyle=numeric-comp, maxbibnames=99, sorting=nyt,
			sortcites=true, firstinits=true, sorting=none]{biblatex}
\input{bibphys.def}

\usepackage{amsmath,amsfonts,amssymb}

\usepackage{maths_min}

\usepackage[pdftex, colorlinks=true, breaklinks=true,
	urlcolor=blue, linkcolor=blue, citecolor=red, linktocpage,
	pdfcreator={PDFLaTeX}, pdfauthor={Harold Erbin}
]{hyperref}
\usepackage[all]{hypcap}

\hypersetup{
	pdftitle={Five-dimensional Janis-Newman algorithm},
}

\numberwithin{equation}{section}

\newcommand{\email}[1]{\thanks{\href{mailto:#1}{\texttt{#1}}}}

\newcommand{\preprint}[0]{CPHT–RR086.1014}
\fancypagestyle{preprint}{
	\fancyhf{}

	\fancyhead[R]{\textsf{\preprint}}
}

\title{Five-dimensional Janis–Newman algorithm}

\author[1]{Harold Erbin\email{erbin@lpthe.jussieu.fr}}
\author[2]{Lucien Heurtier\email{lucien.heurtier@cpht.polytechnique.fr}}

\affil[1]{Sorbonne Universités, UPMC Univ Paris 06, UMR 7589, LPTHE, F-75005, Paris, France}
\affil[1]{CNRS, UMR 7589, LPTHE, F-75005, Paris, France}
\affil[2]{CPHT, École Polytechnique, CNRS, 91128 Palaiseau, France}

\begin{document}

\maketitle
\thispagestyle{preprint}

\begin{abstract} 
The Janis–Newman algorithm has been shown to be successful in finding new stationary solutions of four-dimensional gravity.
Attempts for a generalization to higher dimensions have already been found for the restricted cases with only one angular momentum.
In this paper we propose an extension of this algorithm to five dimensions with two angular momenta – using the prescription of G. Giampieri – through two specific examples, that are the Myers–Perry and BMPV black holes.
We also discuss possible enlargements of our prescriptions to other dimensions and maximal number of angular momenta, and show how dimensions higher than six appear to be much more challenging to treat within this framework.
Nonetheless this general algorithm provides a unification of the formulation in $d = 3, 4, 5$ of the Janis–Newman algorithm, from which expose several examples including the BTZ black hole.
\end{abstract}

\tableofcontents

\section{Introduction}

Exact solutions of Einstein gravity coupled to matter and gauge fields are of great interest because they allow to test and to strengthen our knowledge of the theory.
In particular rotating metrics have many applications and display interesting structures, but they are also very challenging to discover.
The problem has been greatly studied in all dimensions.
While in four dimensions we have at our disposal many theorems on the classification of solutions, this is not the case for higher dimensions and the bestiary for solutions is much wider and less understood~\cite{emparan_black_2008, adamo_kerr-newman_2014}.
In particular important solutions have not yet been discovered, such as charged rotating black holes with several angular momenta (in pure Einstein–Maxwell gravity).

A detour to solving exactly Einstein equations relies on solution generating algorithms, producing new (rotating) possible candidates from a well known (static) solutions, via a step-by-step procedure of transformation.
Such techniques, even if they do not always provide full solutions, may give an ansatz and interesting insights on the structure of the solutions.

A proposal for such an algorithm has been explained in 1965 by Janis and Newman (JN)~\cite{newman_note_1965}, leading to the discovery of the Kerr–Newman metric as a first success~\cite{newman_metric_1965, adamo_kerr-newman_2014}.
It relies on performing a complex change of coordinates on the metric written in a null tetrad basis.
In this scenario, the metric is written using – arbitrarily chosen – real functions but coordinates are allowed to transform in the complex plane, the only condition on the function being that they reduce to the static functions when the coordinates are restricted to the real axis.
As a consequence of this arbitrary choice, the algorithm presents an ambiguity concerning the complexification of the metric components.
Nonetheless many solutions have been found in four dimensions~\cite{herrera_complexification_1982, drake_application_1997, drake_uniqueness_2000, yazadjiev_newman-janis_2000, ibohal_rotating_2005}, and a set of rules has been established.

Another formulation of this algorithm has been proposed by Giampieri~\cite{giampieri_introducing_1990, erbin_janis-newman_2015} and it provides a way of performing the algorithm much more simply than in the initial formulation due to Newman and Janis.
Indeed one does not have to invert the metrics nor to find a null tetrad basis in order to apply the complex change of coordinates.
This is particularly useful when working in an arbitrary number of dimensions.
On the other hand the replacement of the functions inside the metric is the same as in the original prescription.
More specifically our results could also be apply to the search of new rotating solutions in (gauged) supergravity.

The JN algorithm has been applied in higher dimensions by Xu who added one angular momentum to Schwarzschild–Tangherlini solution~\cite{xu_exact_1988}.
The goal of our paper is to show how to add all the angular momenta allowed by the spacetime isometries.
In particular we focus on five dimensions and we detail two examples: the Myers–Perry black hole~\cite{myers_black_1986} and the Breckenridge–Myers–Peet–Vafa (BMPV) extremal black hole~\cite{breckenridge_d-branes_1997}.
We show that parametrizing the metric on the sphere by direction cosines was a key step to generalize the transformation to any dimension since these coordinates are totally symmetric under interchange of angular momenta, which is not the case for spherical coordinates.
We show that we can perform an extended version of the JN algorithm to recover both solutions, starting with a static metric.

While it is possible to obtain the correct structure of the metric in any dimension (for Myers–Perry-like metrics), we show in appendix that it is very challenging to determine the functions inside the metric.

As a refreshing aside, we point out that while Kim had shown that the rotating BTZ black hole~\cite{banados_black_1992} can be obtained from the JN algorithm~\cite{kim_notes_1997, kim_spinning_1999}, our approach includes also this $3$-dimensional example and allows us to recover his result more directly.

The paper is organized as follows.
In section~\ref{sec:introduction-jna} we recall the JN algorithm in both its original version and Giampieri's effective formulation, and we illustrate them on the Kerr–Newman black hole.
In section~\ref{sec:MP-5d} we recover Myers–Perry black hole in five dimensions with two angular momenta, while in section~\ref{sec:BMPV} we derive the BMPV black hole.
In appendix~\ref{app:coord} we gather formula on coordinate systems in $d$ dimensions, and then in $4$ and $5$ dimensions.
Appendix~\ref{app:higher-dim-jna} presents the derivation of the rotating metric in any dimensions together with three examples (flat space, Myers–Perry with one angular momentum and BTZ black hole).
Finally in the appendix~\ref{app:bmpv-second-approach} we give another derivation of the BMPV black hole.

\section{From Janis–Newman procedure to Giampieri's effective approach}
\label{sec:introduction-jna}

\subsection{Original formulation}

In this section we outline the original JN algorithm~\cite{newman_note_1965, newman_metric_1965}, while more detailed reviews can be found in~\cite{drake_uniqueness_2000, whisker_braneworld_2008, keane_extension_2014}.
This algorithm provides a way to derive rotational solutions in four dimensions, applying peculiar complex transformations to a static metric.
The simple case where there is only one function appearing in the metric as well as more general cases  is illustrated in the previous references. We will restrict our study to the case involving only one function of $r$ in what follows for simplifying the extension of the algorithm to higher dimensions.\footnote{Note that a more general formulation of static metrics could in principle be used, e.g. \cite{ibohal_rotating_2005, yazadjiev_newman-janis_2000}, but would be more subtle to extend to higher dimensions.}

Using a generic function $f(r)$ one can start with an initial (static) metric of the form
\begin{equation}
	\label{metric:spherical-f}
	\dd s^2 = - f(r)\, \dd t^2 + f(r)^{-1} \dd r^2 + r^2 \dd \Omega^2, \qquad
	\dd \Omega^2 = \dd\theta^2 + \sin^2 \theta\; \dd \phi^2\,.
\end{equation}

The algorithm proceeds as follows:
\begin{enumerate}
	\item {\bf Transformation to null coordinates}
	
	One first defines the null direction
	\begin{equation}
		\label{change:tr-ur}
		\dd u = \dd t - f^{-1} \dd r
	\end{equation} 
	and transforms the metric to obtain
	\begin{equation}
		\label{metric:spherical-null}
		\dd s^2 = - f\, \dd u^2 - 2\, \dd u \dd r + r^2\, \dd \Omega^2.
	\end{equation} 
	
	\item {\bf Introduction of the Newman–Penrose null tetrads formalism}
	
	One introduces a set of null tetrads
	\begin{equation}
		Z^\mu_a = \{\ell^\mu, n^\mu, m^\mu, \bar m^\mu\}
	\end{equation}
	and rewrite the inverse metric under the form
	\begin{equation}
		g^{\mu\nu} = \eta^{ab} Z^\mu_a Z^\nu_b
	\end{equation} 
	with
	\begin{equation}
		\eta^{ab} =
		\begin{pmatrix}
			0 & -1 & 0 & 0 \\
			-1 & 0 & 0 & 0 \\
			0 &  0 & 0 & 1 \\
			0 &  0 & 1 & 0 \\
		\end{pmatrix}.
	\end{equation} 
	
	For the metric \eqref{metric:spherical-null} the tetrads are found to be
	\begin{equation}
		\ell^\mu = \delta_r^\mu, \qquad
		n^\mu = \delta_u^\mu -\frac{f}{2}\; \delta_r^\mu, \qquad
		m^\mu = \frac{1}{\sqrt{2} \bar r} \left(\delta_\theta^\mu + \frac{i}{\sin \theta} \delta_\phi^\mu \right).
	\end{equation} 
	At this point $r \in \R$ such that $\bar r = r$.
	
	\item {\bf Complexification of two coordinates} 
	
	Here comes the major idea of the algorithm.
	One can allow the coordinates $u$ and $r$ to take complex values but under three restrictions:
	\begin{itemize}
		\item $\ell^\mu$ and $n^\mu$ must still be real;
		\item $m^\mu$ and $\bar m^\mu$ must still be complex conjugated to each other;
		\item one should recover the previous basis for $r \in \mathbb R$.
	\end{itemize}
	Consequently, the function $f(r)$ has to be transformed into a new function $\tilde f(r, \bar r) \in \mathbb R$ such that $\tilde f(r, r) = f(r)$.
	This point is the most difficult to achieve since there is no unique rule to choose a particular complexification and one would always need to check whether any choice of complexification effectively leads to a solution obeying Einstein equations.
	
	\item {\bf Complex transformation}
	
	Once one has opened the possibility to have complex coordinates, it becomes allowed to carry out a complex change of coordinates on $u$ and $r$~\footnote{This transformation can be made more general~\cite{demianski_new_1972, drake_uniqueness_2000, whisker_braneworld_2008}.}
	\begin{equation}
		\label{change:complexification-ur}
		u = u' + ia \cos \theta, \qquad
		r = r' - ia \cos \theta, \qquad
		\theta' = \theta, \qquad
		\phi' = \phi,
	\end{equation} 
	where $a$ is an arbitrary real parameter, with the restriction that $r', u' \in \mathbb R$ and one has $\tilde f(r, \bar r) = \tilde f(r', \theta')$.
	Under this transformation the tetrads turn into
	\begin{equation}
	\begin{gathered}
		\label{eq:tetrads}
		\ell'^\mu = \delta_r^\mu, \qquad
		n'^\mu = \delta_u^\mu -\frac{f}{2}\; \delta_r^\mu, \\
		m'^\mu = \frac{1}{\sqrt{2} (r + ia \cos \theta)} \left(
			\delta_\theta^\mu
			+ \frac{i}{\sin \theta} \delta_\phi^\mu
			- ia \sin \theta\, (\delta_u^\mu - \delta_r^\mu) \right).
	\end{gathered}
	\end{equation} 
	
	\item {\bf Reconstitution of the metric}
	
	One can finally rebuild the metric $g'^{\mu\nu}$ from the new tetrad $\{\ell'^\mu, n'^\mu, m'^\mu, \bar m'^\mu\}$, find the covariant components $g'_{\mu\nu}$ and eventually operate a change of coordinate to interpret its geometry.

\end{enumerate}

As we will see, this algorithm can be applied to the standard Reissner–Nordström solution, generating after deriving all those steps, the Kerr–Newman rotating solution.

\subsection{Giampieri's shortcut}
\label{sec:jna:shortcut}

Finding the contravariant components of the metric is easy for the example given above, but it may become hard when one increases the complexity of the metric, or the number of spacetime dimensions.
In an unpublished paper~\cite{giampieri_introducing_1990}, Giampieri suggested a new prescription to avoid such laborious computations.
It consists in complexifying directly the metric \eqref{metric:spherical-null}, doing the change of coordinates and finally removing the complex contribution by using a specific ansatz.
This approach has been reviewed in~\cite{erbin_janis-newman_2015}.

The algorithm can be formulated as follows:
\begin{enumerate}
	\item {\bf Transformation to null coordinates}
	
	As in step 1) of the JN algorithm, start writing the metric using the null coordinate $u$
	\begin{equation}
		\label{metric:spherical:null}
		\dd s^2 = - f\, \dd u^2 - 2\, \dd u \dd r + r^2\, \dd \Omega^2.
	\end{equation} 
	
	\item {\bf Complexification}
	
	Coordinates $u$ and $r$ are directly let to become complex and the metric \eqref{metric:spherical:null} is complexified.
	\begin{equation}
		\dd s^2 = - \tilde f\, \dd u^2 - 2\, \dd u \dd r + \abs{r}^2 \dd \Omega^2,
	\end{equation} 
	where again $\tilde f = \tilde f(r, \bar r)$ is the real-valued function resulting from $f$.
	At this step the metric continues to be real.
	
	\item {\bf Complex transformation}
	
	Introducing a new angle $\chi$, apply the change of coordinates \eqref{change:complexification-ur}
	\begin{equation}
		\label{change:jna-ur}
		u = u' + ia \cos \chi, \qquad
		r = r' - ia \cos \chi, \qquad
		\theta' = \theta, \qquad
		\phi' = \phi,
	\end{equation} 
	and
	\begin{equation}
		\label{change:jna-diff-ur-old}
		\dd u = \dd u' - ia \sin \chi\; \dd \chi, \qquad
		\dd r = \dd r' + ia \sin \chi\; \dd \chi.
	\end{equation}
	
	This step can be interpreted as an embedding in a $5$-dimensional theory from which the final metric will be extracted.
	
	The solution becomes (omitting the primes)
	\begin{equation}
		\dd s'^2 = - \tilde f (\dd u - ia \sin\chi\, \dd\chi)^2
			- 2\, (\dd u - ia \sin\chi\, \dd\chi) (\dd r + ia \sin\chi\, \dd\chi)
			+ \rho^2\, \dd \Omega^2
	\end{equation}
	where the variable $\rho^2$ has been defined as
	\begin{equation}
		\label{metric:rotating:rho}
		\rho^2 = r^2 + a^2 \cos^2 \theta.
	\end{equation} 
	
	\item {\bf Slice fixing}
	
	As explained one should fix the angle $\chi$ to recover a $4$-dimensional solution.
	The correct way is to make the ansatz
	\begin{subequations}
	\label{eq:giampieri-ansatz}
	\begin{equation}
		i\, \dd \chi = \sin \chi\, \dd \phi
	\end{equation}
	followed by the replacement
	\begin{equation}
		\chi = \theta.
	\end{equation}
	\end{subequations}
	One gets
	\begin{equation}
		\label{metric:rotating:Kerr-coord}
		\dd s^2 = - \tilde f\, (\dd u - a \sin^2 \theta\, \dd\phi)^2
			- 2\, (\dd u - a \sin^2 \theta\, \dd\phi) (\dd r + a \sin^2 \theta\, \dd\phi)
			+ \rho^2 \dd \Omega^2.
	\end{equation}
	
	\item {\bf Standard coordinate redefinition}
	
	Using
	\begin{equation}
		\label{change:rotating:g-h}
		\dd u = \dd t' - g(r) \dd r, \qquad
		\dd \phi = \dd \phi' - h(r) \dd r.
	\end{equation} 
	one can go into Boyer–Lindquist coordinates by imposing the conditions $g_{tr} = g_{r\phi'} = 0$, which are solved for
	\begin{equation}
		\label{change:rotating:ur-bl}
		g = \frac{r^2 + a^2}{\Delta}, \qquad
		h = \frac{a}{\Delta},
	\end{equation} 
	where $\Delta$ is defined as
	\begin{equation}
		\label{metric:rotating:delta}
		\Delta = \tilde f \rho^2 + a^2 \sin^2 \theta.
	\end{equation} 
	Obviously this change of variable makes sense only if $g$ and $h$ are functions of $r$ only.
	One thus gets the metric (omitting the primes)~\cite{visser_kerr_2007}
	\begin{equation}
		\label{metric:rotating:bl-coord}
		\dd s^2 = - \tilde f\, \dd t^2
			+ \frac{\rho^2}{\Delta}\, \dd r^2
			+ \rho^2 \dd\theta^2
			+ \frac{\Sigma^2}{\rho^2} \sin^2 \theta\; \dd\phi^2
			+ 2a (\tilde f - 1) \sin^2 \theta\; \dd t \dd\phi
	\end{equation} 
	with
	\begin{equation}
		\frac{\Sigma^2}{\rho^2} = r^2 + a^2 + a g_{t\phi} \,.
	\end{equation} 
	
\end{enumerate}

Giampieri's prescription may seem to be less rigorous than the original formulation, but one may trace the ansatz \eqref{eq:giampieri-ansatz} and its consequences to the tetrad transformation, ensuring that both approaches are totally equivalent.
The fact that this procedure is completely effective allows us to extend the algorithm to higher dimensions as we will illustrate in sections~\ref{sec:MP-5d} and~\ref{sec:BMPV}.

Another peculiar feature of this approach is that one should consider the complexification of the differentials and the complexification of the functions appearing inside the metric as two different processes: one can derive general formula as we did by taking $f$ arbitrary while the differentials are transformed.
From this point of view the $r^2$ factor in front of $\dd\Omega^2$ can also be considered as a function with its own complexification.

\subsection{Kerr–Newman}
\label{sec:Kerr-Newman}

As mentioned previously, the generic function $f(r)$ can be taken to be the one of the Schwarzschild black hole solution to generate the Kerr solution.
More generally, one can even add an electric charge and start from the Reissner–Nordström black hole to end up with the Kerr–Newman solution~\cite{newman_metric_1965}.
The initial metric is taken to be
\begin{equation}
	\ s^2 = - f(r)\, \dd t^2 + f(r)^{-1} \dd r^2 + r^2 \dd \Omega^2, \qquad
	f(r) = 1 - \frac{2 m}{r} + \frac{q^2}{r^2}.
\end{equation} 
The electromagnetic field is encoded by the gauge field
\begin{equation}
	\label{pot:kerr-newman}
	A = \frac{q}{r}\; \dd t.
\end{equation} 
The function $f$ can be complexified as
\[
	\tilde f(r) = 1 - m \left(\frac{1}{r} + \frac{1}{\bar r}\right) + \frac{q^2}{\abs{r}^2}
		= 1 - m\, \frac{r + \bar r}{\abs{r}^2} + \frac{q^2}{\abs{r}^2}
\]
giving
\begin{equation}
	\tilde f(r) = 1 + \frac{q^2 - 2m r}{\rho^2}
\end{equation}
where we recall that $\rho^2 = r^2 + a^2 \cos^2 \theta$.

As it is shown in ~\cite{newman_metric_1965, drake_uniqueness_2000}, inserting this function into \eqref{metric:rotating:bl-coord} provides the Kerr–Newman metric
\begin{equation}
	\dd s^2 = - \tilde f\, \dd t^2
		+ \frac{\rho^2}{\Delta} \dd r^2
		+ \rho^2 \dd\theta^2
		+ \frac{\Sigma^2}{\rho^2} \sin^2 \theta\; \dd\phi^2
		+ 2a (\tilde f - 1) \sin^2 \theta\; \dd t \dd\phi\,,
\end{equation}
where the quantities $\Delta$ and $\Sigma$ are defined by
\begin{subequations}
\begin{gather}
	\frac{\Sigma^2}{\rho^2} = r^2 + a^2 - \frac{q^2 - 2m r}{\rho^2}\, a^2 \sin^2 \theta \,, \\
	\Delta = r^2 - 2m r + a^2 + q^2.
\end{gather}
\end{subequations}
Here $\Delta$ depends only on $r$ and the transformation \eqref{change:rotating:ur-bl} to Boyer–Lindquist coordinates is well defined.

As far as the gauge field is concerned, it has been shown in~\cite{erbin_janis-newman_2015} how to build its rotating equivalent with both the original JN algorithm and Giampieri's prescriptions (see also~\cite{keane_extension_2014, adamo_kerr-newman_2014} for another approach).
The latter procedure providing an effective – but cleaner – computation, will keep being used in what follows.

At first, taking the gauge field \eqref{pot:kerr-newman} of the Reissner–Nordström solution and translating it into the $(u, r)$ set of coordinates gives
\begin{equation}
	A = \frac{q}{r}\, (\dd u + f^{-1} \dd r).
\end{equation} 
which can be reduced to
\begin{equation}
	A = \frac{q}{r}\; \dd u
\end{equation} 
thanks to a simple gauge transformation.

Using again the rule
\begin{equation}
	\frac{1}{r} \longrightarrow \frac{1}{2} \left(\frac{1}{r} + \frac{1}{\bar r}\right)
		= \frac{r}{\rho^2}\,,
\end{equation} 
and the transformations \eqref{change:jna-ur}, gives
\begin{equation}
	A = \frac{q r}{\rho^2}\, (\dd u - a \sin^2 \theta\, \dd \phi).
\end{equation} 

One can finally go to Boyer–Lindquist coordinates, while applying another gauge transformation to remove the $A_r$ contribution (independent of $r$), which provides the standard form of the Kerr–New\-man gauge field
\begin{equation}
	A = \frac{q r}{\rho^2}\, (\dd t - a \sin^2 \theta\, \dd \phi).
\end{equation} 

In what follows we propose a generalization of this procedure by increasing the number of dimensions and momenta.

\section{Myers–Perry solution in five dimensions}
\label{sec:MP-5d}

In this section we show how to recover Myers–Perry black hole in five dimensions through Giampieri's prescription.
This is a solution of $5$-dimensional pure Einstein theory which possesses two angular momenta and it generalizes the Kerr black hole.
The importance of this solution lies in the fact that it can be constructed in any dimension.

Let us start with the five-dimensional Schwarzschild–Tangherlini metric
\begin{equation}
	\dd s^2 = - f(r)\, \dd t^2 + f(r)^{-1}\, \dd r^2 + r^2\, \dd \Omega_3^2
\end{equation}
where $\dd \Omega_3^2$ is the metric on $S^3$, which can be expressed in Hopf coordinates (see appendix~\ref{app:coord:5d:hopf})
\begin{equation}\label{eq:coord-S3-spherical}
	\dd \Omega_3^2 = \dd\theta^2 + \sin^2 \theta\, \dd\phi^2 + \cos^2 \theta\, \dd\psi^2\,,
\end{equation} 
and the function $f(r)$ is given by
\begin{equation}
	f(r) = 1 - \frac{m}{r^2}\,.
\end{equation}

An important feature of the JN algorithm is the fact that a given set of transformations in the $(r,\phi)$-plane generates rotation in the latter.
Generating several angular momenta in different 2-planes would then require successive applications of the JN algorithm on different hypersurfaces.
In order to do so, one has to identify what are the 2-planes which will be submitted to the algorithm.
In five dimensions, the two different planes that can be made rotating are the planes $(r,\phi)$ and $(r,\psi)$.
We claim that it is necessary to dissociate the radii of these 2-planes in order to apply separately the JN algorithm on each plane and hence to generate two distinct angular momenta.
In order to dissociate the parts of the metric that correspond to the rotating and non-rotating $2$-planes, one can protect the function $r^2$ to be transformed under complex transformations in the part of the metric defining the plane which will stay static.
We thus introduce the function
\begin{equation}
	R(r) = \Re(r)
\end{equation} 
such that the metric in null coordinates reads
\begin{equation}
	\label{5d-jna:metric:static:general-null}
	\dd s^2 = - \dd u\, (\dd u + 2 \dd r)
		+ (1 - f)\, \dd u^2
		+ r^2 (\dd\theta^2 + \sin^2 \theta\, \dd\phi^2) + R^2 \cos^2 \theta\, \dd\psi^2.
\end{equation} 
The first transformation – hence concerning the $(r,\phi)$-plane – is
\begin{equation}
	\label{eq:5d-ansatz-hopf-1}
	\begin{gathered}
		u = u' + i a \cos \chi_1, \qquad
		r = r' - i a \cos \chi_1, \\
		i\, \dd \chi_1 = \sin \chi_1\ \dd\phi, \qquad\text{~~with~~}\chi_1 = \theta, \\
		\dd u = \dd u' - a \sin^2 \theta\, \dd\phi, \qquad
		\dd r = \dd r' + a \sin^2 \theta\, \dd\phi,
	\end{gathered}
\end{equation}
and $f$ is replaced by $\tilde f^{\{1\}} = \tilde f^{\{1\}}(r, \theta)$.
Indeed we need to keep track of the order of the transformation, since the function $f$ will be complexified twice consecutively.
On the other hand $R(r) = \Re(r)$ transforms into $R(r) = r'$ and we find (omitting the primes)
\begin{equation}
	\begin{aligned}
	\dd s^2 = - \dd u^2 &- 2\, \dd u \dd r
		+ \big(1 - \tilde f^{\{1\}} \big) (\dd u - a \sin^2 \theta\, \dd \phi)^2
		+ 2 a \sin^2 \theta\, \dd r \dd \phi \\
		&+ (r^2 + a^2 \cos^2 \theta) \dd\theta^2
		+ (r^2 + a^2) \sin^2 \theta\, \dd\phi^2
		+ r^2 \cos^2 \theta\, \dd \psi^2.
	\end{aligned}
\end{equation} 
The function $\tilde f^{\{1\}}$ is
\begin{equation}
	\tilde f^{\{1\}} = 1 - \frac{m}{\abs{r}^2} = 1 - \frac{m}{r^2 + a^2 \cos^2 \theta}.
\end{equation} 

In addition to the terms present in \eqref{5d-jna:metric:static:general-null} we obtain new components corresponding to the rotation of the first plane $(r,\phi)$.
We find the same terms as in \eqref{5d-jna:metric:static:general-null} plus other terms that corresponds to the rotation in the first plane.
Transforming now the second one – $(r, \psi)$ – the transformation is~\footnotemark
\footnotetext{The easiest justification for choosing the sinus here is by looking at the transformation in terms of direction cosines, see appendix~\ref{sec:higher-dim:examples:MP-5d}.
Otherwise this term can be guessed by looking at Myers–Perry non-diagonal terms.}
\begin{equation}
	\label{eq:5d-ansatz-hopf-2}
	\begin{gathered}
		u = u' + i b\, \sin \chi_2, \qquad
		r = r' - i b\, \sin \chi_2, \\
		i\, \dd \chi_2 = - \cos \chi_2\, \dd\psi, \qquad \text{~~with~~}\chi_2 = \theta, \\
		\dd u = \dd u' - b \cos^2 \theta\, \dd\psi, \qquad
		\dd r = \dd r' + b \cos^2 \theta\, \dd\psi,
	\end{gathered}
\end{equation}
can be applied directly to the metric
\begin{equation}
	\begin{aligned}
	\dd s^2 = - \dd u^2 &- 2\, \dd u \dd r
		+ \big(1 - \tilde f^{\{1\}} \big) (\dd u - a \sin^2 \theta\, \dd \phi)^2
		+ 2 a \sin^2 \theta\, \dd R \dd \phi \\
		&+ \rho^2 \dd\theta^2
		+ (R^2 + a^2) \sin^2 \theta\, \dd\phi^2
		+ r^2 \cos^2 \theta\, \dd \psi^2
	\end{aligned}
\end{equation} 
where we introduced once again the function $R(r) = \Re(r)$ to protect the geometry of the first plane to be transformed under complex transformations.

The final result (using again $R = r'$ and omitting the primes) becomes
\begin{equation}
	\begin{aligned}
	\dd s^2 = - \dd u^2 &- 2\, \dd u \dd r
		+ \big(1 - \tilde f^{\{1, 2\}} \big) (\dd u - a \sin^2 \theta\, \dd \phi - b \cos^2 \theta\, \dd \psi)^2
		\\
		&+ 2 a \sin^2 \theta\, \dd r \dd \phi
		+ 2 b \cos^2 \theta\, \dd r\dd \psi \\
		&+ \rho^2 \dd\theta^2
		+ (r^2 + a^2) \sin^2 \theta\, \dd\phi^2
		+ (r^2 + b^2) \cos^2 \theta\, \dd \psi^2
	\end{aligned}
\end{equation} 
where
\begin{equation}
	\rho^2 = r^2 + a^2 \cos^2 \theta + b^2 \sin^2 \theta.
\end{equation} 
Furthermore, the function $\tilde f^{\{1\}}$ has been complexified as
\begin{equation}
 	\tilde f^{\{1,2\}} = 1 - \frac{m}{\abs{r}^2 + a^2 \cos^2 \theta}
		= 1 - \frac{m}{r'^2 + a^2 \cos^2 \theta + b^2 \sin^2 \theta}
		= 1 - \frac{m}{\rho^2}.
\end{equation}

The metric can then be transformed into the Boyer–Lindquist (BL) using
\begin{equation}
	\dd u = \dd t - g(r)\, \dd r, \qquad
	\dd\phi = \dd\phi' - h_\phi(r)\, \dd r, \qquad
	\dd\psi = \dd\psi' - h_\psi(r)\, \dd r.
\end{equation} 
Defining the parameters~\footnotemark
\footnotetext{See \eqref{metric:rotating-result-jna-bl-parameters} for a definition of  $\Delta$ in terms of $\tilde f$.}
\begin{equation}
	\Pi = (r^2 + a^2) (r^2 + b^2), \qquad
	\Delta = r^4 + r^2 (a^2 + b^2- m) + a^2 b^2\,,
\end{equation}
the functions can be written
\begin{equation}
	g(r) = \frac{\Pi}{\Delta}, \qquad
	h_\phi(r) = \frac{\Pi}{\Delta}\, \frac{a}{r^2 + a^2}, \qquad
	h_\psi(r) = \frac{\Pi}{\Delta}\, \frac{b}{r^2 + b^2}\,.
\end{equation} 
We get the final metric
\begin{equation}
	\label{metric:rotating-5d-2-moments-bl}
	\begin{aligned}
		\dd s^2 = - \dd t^2
			&+ \big(1 - \tilde f^{\{1, 2\}} \big) (\dd t - a \sin^2 \theta\, \dd \phi - b \cos^2 \theta\, \dd \psi)^2
			+ \frac{r^2 \rho^2}{\Delta}\; \dd r^2 \\
			&+ \rho^2 \dd\theta^2
			+ (r^2 + a^2) \sin^2 \theta\, \dd\phi^2
			+ (r^2 + b^2) \cos^2 \theta\, \dd \psi^2\,.
	\end{aligned}
\end{equation} 

One recovers here the five dimensional Myers–Perry black hole with two angular momenta~\cite{myers_black_1986}.

It is important to mention that following the same prescription in dimensions higher than five does not lead as nicely as we did in five dimensions to the exact Myers–Perry solution.
Indeed we show in appendix~\ref{app:higher-dim-jna} that the transformation of the metric can be done along the same line but that the only – major – obstacle comes from the function $f$ that can not be complexified as expected.
Finding the correct complexification seems very challenging and it may be necessary to use a different complex coordinate transformation in order to perform a completely general transformation in any dimension.
It might be possible to gain insight into this problem by computing the transformation within the framework of the tetrad formalism.

One may think that a possible solution would be to replace complex numbers by quaternions, assigning one angular momentum to each complex direction but it is straightforward to check that this approach is not working.

\section{BMPV black hole}
\label{sec:BMPV}

\subsection{Few properties and seed metric}

In this section we focus on another example in five dimensions, which is the BMPV black hole~\cite{breckenridge_d-branes_1997, gauntlett_black_1999}.
This solution possesses many interesting properties, in particular it can be proven that it is the only rotating BPS asymptotically flat black hole in five dimensions with the corresponding near-horizon geometry~\footnotemark{}~\cites[sec.~7.2.2, 8.5]{emparan_black_2008}{reall_higher_2003}.
\footnotetext{Other possible near-horizon geometries are $S^1 \times S^2$ (for black rings) and $T^3$, even if the latter does not seem really physical.
BMPV horizon corresponds to the squashed $S^3$.}
It is interesting to notice that even if this extremal solution is a slowly rotating metric, it is an exact solution (whereas Einstein equations need to be truncated for consistency of usual slow rotation).

For a rotating black hole the BPS and extremal limits do not coincide~\cites[sec.~7.2]{emparan_black_2008}[sec.~1]{gauntlett_black_1999}: the first implies that the mass and the electric charge are equal~\footnote{It is a consequence from the BPS bound $m \ge \sqrt{3}/2\, \abs{q}$.}, while extremality~\footnote{Regularity is given by a bound, which is saturated for extremal black holes.} implies that one linear combination of the angular momenta vanishes, and for this reason we set $a = b$ from the beginning~\footnote{If we had kept $a \neq b$ we would have discovered later that one cannot transform the metric to Boyer–Lindquist coordinates without setting $a = b$.}.
We are thus left with two parameters that we take to be the mass and one angular momentum.

In the non-rotating limit BMPV black hole reduces to the charged extremal Schwarzschild--Tangherlini (with equal mass and charge) written in isotropic coordinates.
For non-rotating black hole the extremal and BPS limit are equivalent.

Both the charged extremal Schwarzschild–Tangherlini and BMPV black holes are solutions of minimal $d = 5$ supergravity (Einstein–Maxwell plus Chern–Simons) whose action is~\cites[sec.~1]{gauntlett_black_1999}[sec.~2]{aliev_superradiance_2014}[sec.~2]{gauntlett_all_2003}
\begin{equation}
	S = - \frac{1}{16\pi G} \int \left(R\; \hodge{1} + F \wedge \hodge{F} + \frac{2\lambda}{3 \sqrt{3}}\, F \wedge F \wedge A \right),
\end{equation} 
where supersymmetry is imposing $\lambda = 1$.

Since extremal limits are different for static and rotating black holes we can guess that the black hole we will obtain from the algorithm will not be a solution of the equations of motion and we will need to take some limit.

The charged extremal Schwarzschild–Tangherlini black hole is taken as a seed metric~\cites[sec.~3.2]{gauntlett_all_2003}[sec.~4]{gibbons_supersymmetric_1994}[sec.~3]{tangherlini_schwarzschild_1963}
\begin{equation}
	\label{metric:5d-bmpv:seed-metric}
	\dd s^2 = - H^{-2} \dd t^2 + H (\dd r^2 + r^2 \dd\Omega_3^2 )
\end{equation} 
where $\dd\Omega_3^2$ is the metric of the $3$-sphere written in
\eqref{eq:coord-S3-spherical}.
The function $H$ is harmonic
\begin{equation}
	H(r) = 1 + \frac{m}{r^2},
\end{equation} 
and the electromagnetic field reads
\begin{equation}
	\label{eq:5d-bmpv:seed-em-field}
	A = \frac{\sqrt{3}}{2 \lambda}\; \frac{m}{r^2}\; \dd t = (H - 1)\, \dd t.
\end{equation} 

In the next subsections we apply successively the transformations \eqref{eq:5d-ansatz-hopf-1} and \eqref{eq:5d-ansatz-hopf-2} with $a = b$ in the case $\lambda = 1$ because we are searching a supersymmetric solution.

\subsection{Transforming the metric}

The transformation to $(u, r)$ coordinates of the seed metric \eqref{metric:5d-bmpv:seed-metric}
\begin{equation}
	\dd t = \dd u + H^{3/2}\, \dd r
\end{equation} 
gives
\begin{equation}
	\dd s^2 = - H^{-2} \dd u^2 - 2 H^{-1/2} \dd u \dd r + H r^2 \dd\Omega_3^2
		= - H^{-2} \big(\dd u - 2 H^{3/2} \dd r \big)\, \dd u + H r^2 \dd\Omega_3^2.
\end{equation} 

For transforming the above metric one should follow the recipe of the previous section: transformations \eqref{eq:5d-ansatz-hopf-1}
\begin{equation}
	u = u' + i a \cos \theta, \qquad
	\dd u = \dd u' - a \sin^2 \theta\, \dd\phi,
\end{equation}
and \eqref{eq:5d-ansatz-hopf-2}
\begin{equation}
	u = u' + i a\, \sin \theta, \qquad
	\dd u = \dd u' - a \cos^2 \theta\, \dd\psi
\end{equation} 
are performed one after another, transforming each time only the terms that pertain to the corresponding rotation plane~\footnotemark.
In order to preserve the isotropic form of the metric the function $H$ is complexified everywhere (even when it multiplies terms that belong to the other plane).
\footnotetext{For another approach see appendix~\ref{app:bmpv-second-approach}.}

Since the procedure is exactly similar to the Myers–Perry case we give only the final result in $(u, r)$ coordinates
\begin{equation}
	\label{metric:5d-bmpv:bmpv-metric-ur-before-limit}
	\begin{aligned}
		\dd s^2 = &- \tilde H^{-2} \big(\dd u
				- a (1 - \tilde H^{3/2}) (\sin^2 \theta\, \dd\phi + \cos^2 \theta\, \dd\psi) \big)^2 \\
			&- 2 \tilde H^{-1/2} \big(\dd u - a (1 - \tilde H^{3/2})\, (\sin^2 \theta\, \dd\phi + \cos^2 \theta\, \dd\psi) \big)\, \dd r \\
			&+ 2 a \tilde H\, (\sin^2 \theta\, \dd\phi + \cos^2 \theta\, \dd\psi)\, \dd r
			- 2 a^2 \tilde H \cos^2 \theta \sin^2 \theta\, \dd\phi \dd\psi
			\\
			&+ \tilde H\, \Big(
				(r^2 + a^2) (\dd \theta^2 + \sin^2 \theta\, \dd\phi^2 + \cos^2 \theta\, \dd\psi^2)
				+ a^2 (\sin^2 \theta\, \dd\phi + \cos^2 \theta\, \dd\psi)^2 \Big).
	\end{aligned}
\end{equation} 
After both transformations the resulting function $\tilde H$ is
\begin{equation}
	\label{eq:5d-bmpv:tilde-H}
	\tilde H = 1 + \frac{m}{r^2 + a^2 \cos^2\theta + a^2 \sin^2\theta}
		= 1 + \frac{m}{r^2 + a^2}
\end{equation}
which does not depend on $\theta$.

It is easy to check that the Boyer–Lindquist transformation
\begin{equation}
	\dd u = \dd t - g(r)\, \dd r, \qquad
	\dd\phi = \dd\phi' - h_\phi(r)\, \dd r, \qquad
	\dd\psi = \dd\psi' - h_\psi(r)\, \dd r
\end{equation} 
is ill-defined because the functions depend on $\theta$.
The way out is to take the extremal limit alluded above.

Following the prescription of \cite{breckenridge_d-branes_1997, gauntlett_black_1999} and taking the extremal limit
\begin{equation}
	\label{eq:5d-bmpv-extremal-limit}
	a, m \longrightarrow 0, \qquad
	\text{imposing} \qquad
	\frac{m}{a^2} = \cst \,,
\end{equation}
one gets at leading order
\begin{equation}
	\tilde H(r) = 1 + \frac{m}{r^2} = H(r), \qquad
	a\, (1 - \tilde H^{3/2}) = - \frac{3\, m a}{2\, r^2}
\end{equation} 
which translate into the metric
\begin{equation}
	\begin{aligned}
		\dd s^2 = - H^{-2}\, & \left(\dd u
				+ \frac{3\, m a}{2\, r^2}\, (\sin^2 \theta\, \dd\phi + \cos^2 \theta\, \dd\psi) \right)^2 \\
			&- 2 H^{-1/2} \left( \dd u + \frac{3\, m a}{2\, r^2}\, (\sin^2 \theta\, \dd\phi + \cos^2 \theta\, \dd\psi) \right) \dd r \\
			&+ H\, r^2 (\dd \theta^2 + \sin^2 \theta\, \dd\phi^2 + \cos^2 \theta\, \dd\psi^2).
	\end{aligned}
\end{equation} 
Then Boyer–Lindquist functions are
\begin{equation}
	g(r) = H(r)^{3/2}, \qquad
	h_\phi(r) = h_\psi(r) = 0
\end{equation} 
and one gets the metric in $(t, r)$ coordinates
\begin{equation}
	\label{metric:5d-bmpv:bmpv-metric}
	\begin{aligned}
		\dd s^2 = &- \tilde H^{-2} \left(\dd t
				+ \frac{3\, m a}{2\, r^2}\, (\sin^2 \theta\, \dd\phi + \cos^2 \theta\, \dd\psi) \right)^2 \\
			&+ \tilde H\, \Big(\dd r^2 + r^2 \big( \dd \theta^2 + \sin^2 \theta\, \dd\phi^2 + \cos^2 \theta\, \dd\psi^2 \big) \Big).
	\end{aligned}
\end{equation} 
We recognize here the BMPV solution~\cites[p.~4]{breckenridge_d-branes_1997}[p.~16]{gauntlett_black_1999}.
The fact that this solution has only one rotation parameter can be seen more easily in Euler angle coordinates~\cites[sec.~3]{gauntlett_black_1999}[sec.~2]{gibbons_supersymmetric_1999} or by looking at the conserved charges in the $\phi$- and $\psi$-planes~\cite[sec.~3]{breckenridge_d-branes_1997}.

\subsection{Transforming the Maxwell potential}

Following the procedure described in \cite{erbin_janis-newman_2015} and recalled in section~\ref{sec:Kerr-Newman}, one can also derive the gauge field in the rotating framework from the original static one \eqref{eq:5d-bmpv:seed-em-field}.
The latter can be written in the $(u, r)$ coordinates
\begin{equation}
	A = \frac{\sqrt{3}}{2}\, (H - 1)\, \dd u\,,
\end{equation} 
since we can remove the $A_r(r)$ component by a gauge transformation.
One can apply the two JN transformations \eqref{eq:5d-ansatz-hopf-1} and \eqref{eq:5d-ansatz-hopf-2} with $b = a$ to obtain
\begin{equation}
	A = \frac{\sqrt{3}}{2}\, (\tilde H - 1) \Big( \dd u - a\, (\sin^2 \theta\, \dd\phi + \cos^2 \theta\, \dd\psi) \Big).
\end{equation} 

Then going into BL coordinates with \eqref{eq:5d-bl-transformation} provides
\begin{equation}
	A = \frac{\sqrt{3}}{2}\, (\tilde H - 1) \Big( \dd t - a\, (\sin^2 \theta\, \dd\phi + \cos^2 \theta\, \dd\psi) \Big) + A_r(r)\, \dd r.
\end{equation} 
Again $A_r$ depends only on $r$ and can be removed by a gauge transformation.
Applying the extremal limit \eqref{eq:5d-bmpv-extremal-limit} finally gives
\begin{equation}
	A = \frac{\sqrt{3}}{2}\; \frac{m}{r^2} \Big( \dd t - a\, (\sin^2 \theta\, \dd\phi + \cos^2 \theta\, \dd\psi) \Big)\,,
\end{equation}
which is again the result presented in~\cite[p. 5]{breckenridge_d-branes_1997}.

Despite the fact that the seed metric \eqref{metric:5d-bmpv:seed-metric} together with the gauge field \eqref{eq:5d-bmpv:seed-em-field} solves the equations of motion for any value of $\lambda$, the resulting rotating metric solves the equations only for $\lambda = 1$ (see~\cite[sec.~7]{gauntlett_black_1999} for a discussion).
An explanation in this reduction can be found in the limit \eqref{eq:5d-bmpv-extremal-limit} that was needed for transforming the metric to Boyer–Lindquist coordinates and which gives a supersymmetric black hole – which necessarily has $\lambda = 1$.

\subsection{CCLP black hole}

It would be very interesting to find the CCLP black hole~\cite{chong_general_2005}, which is the corresponding non-extremal solution with four independent charges: two angular momenta, an electric charge and the mass.
This black hole is also a solution of $d = 5$ minimal supergravity.

Yet, using our prescription, it appears that the metric of this black hole cannot entirely be recovered.
Indeed all the terms but one are generated by our algorithm, which also provides the correct gauge field.
The missing term is proportional to the electric charge and the current prescription can not generate it.

This issue may be related to the fact that the CCLP solution can not be written as a Kerr–Schild metric but as an extended Kerr–Schild one~\cite{aliev_note_2009, malek_extended_2014}, which includes an additional term proportional to a spacelike vector.
It appears that the missing term corresponds precisely to this additional term in the extended Kerr–Schild metric, and it is well-known that the JN algorithm works mostly for Kerr–Schild metrics.

\section{Conclusion}

In this paper we have presented a generalization of the JN algorithm to five dimensions.
Unlike the previous work from Xu~\cite{xu_exact_1988} which included only one angular momentum, we showed explicitly how to generate two distinct angular momenta on two examples that are the Myers–Perry and BMPV black holes.
The second example shows that the algorithm can be applied to extremal solutions, if one applies appropriate extremal limit to the metric that is obtained.
Furthermore the approach of Giampieri for performing the JN algorithm via an effective procedure seems one more time to be promising for future extensions of solution generation techniques.
The results exposed in this work provide a nice extension of the algorithm for higher dimensional perspectives.

As mentioned in the paper and partially detailed in appendix, the case of the general Myers–Perry solution~\cite{myers_black_1986} as well as the CCLP black hole~\cite{chong_general_2005} are important solutions to be recovered, and it it possible that a generalization of our prescription would be needed.
Another line of work that can be pursued is to apply the algorithm to black rings~\cite{emparan_rotating_2002, emparan_black_2008}.

A major application of our work would be to find the charged solution with two angular momenta of the five-dimensional Einstein–Maxwell.
This problem is highly non-trivial and there is few chances that this technique would work directly~\cite{aliev_rotating_2006}, but one can imagine that a generalization of Demiański's approach~\cite{demianski_new_1972} could lead to new interesting solutions in five dimensions.

Slowly rotating metrics could in principle be derived easily~\cite{aliev_rotating_2006} using our prescriptions and could be a nice playground to understand better higher dimensional solution with $d \ge 6$.
Furthermore the general formalism developed in the appendix appears to be very useful for providing a unified view of the JN algorithm in $d = 3, 4, 5$.

\section*{Acknowledgments}

We wish to thank Nick Halmagyi for interesting discussions.
L. H. would like to thank M. Petropoulos for nice comments and suggestions of reading, as well as the DESY theory group for support during the spring 2014.

For this work, made within the \textsc{Labex Ilp} (reference \textsc{Anr–10–Labx–63}), H. E. was supported by French state funds managed by the \emph{Agence nationale de la recherche}, as part of the programme \emph{Investissements d'Avenir} under the reference \textsc{Anr–11–Idex–0004–02}.

\appendix

\section{Spatial coordinate systems}
\label{app:coord}

This appendix is partly based on~\cite{myers_black_1986, gibbons_general_2005}.
We present formula for any dimension before summarizing them for $4$ and $5$ dimensions.

\subsection{\texorpdfstring{$d$}{d}-dimensional}
\label{app:coord:general-d}

Let's consider $d = N + 1$ dimensional Minkowski space whose metric is denoted by
\begin{equation}
	\dd s^2 = \eta_{\mu\nu}\; \dd x^\mu \dd x^\nu, \qquad
	\mu = 0, \ldots, N.
\end{equation} 
In all the following coordinates systems the time direction can separated from the spatial (positive definite) metric as
\begin{equation}
	\dd s^2 = - \dd t^2 + \dd \Sigma^2, \qquad
	\dd \Sigma^2 = \gamma_{ab}\; \dd x^a \dd x^b, \qquad
	a = 1, \ldots, N,
\end{equation} 
where $x^0 = t$.

We also define
\begin{equation}
	n = \floor{\frac{N}{2}}
\end{equation} 
such that
\begin{equation}
	\label{coord:eq:d-dim-epsilon}
	d + \varepsilon = 2n + 2, \qquad
	N + \varepsilon = 2n + 1, \qquad
	\varepsilon' = 1 - \varepsilon
\end{equation} 
where
\begin{equation}
	\varepsilon = \frac{1}{2} (1 - (-1)^d ) =
	\begin{cases}
		0 & \text{$d$ even (or $N$ odd)} \\
		1 & \text{$d$ odd (or $N$ even)},
	\end{cases}
\end{equation} 
and conversely for $\varepsilon'$.

\subsubsection{Cartesian system}

The usual Cartesian metric is
\begin{equation}
	\dd \Sigma^2 = \delta_{ab} \dd x^a \dd x^b
		= \dd x^a \dd x^a
		= \dd \vec x^2.
\end{equation} 

\subsubsection{Spherical}

Introducing a radial coordinate $r$, the flat space metric can be written as a $(N-1)$-sphere of radius $r$~\cite{tangherlini_schwarzschild_1963}
\begin{equation}
	\label{coord:metric:flat-d:spherical}
	\dd \Sigma^2 = \dd r^2 + r^2 \dd \Omega_{N-1}^2.
\end{equation} 
The term $\dd \Omega_{N-1}^2$ corresponds to the metric on the unit $(N-1)$-sphere $S^{N-1}$, which is parame\-trized by $(N-1)$ angles $\theta_i$ and is defined recursively as
\begin{equation}
	\dd \Omega_{N-1}^2 = \dd \theta_{N-1}^2 + \sin^2 \theta_{N-1} \; \dd \Omega_{N-2}^2.
\end{equation} 

This surface can be embedded in $N$-dimensional flat space with coordinates $X^i$ constrained by
\begin{equation}
	\label{coord:eq:spherical-embedding}
	X^a X^a = 1.
\end{equation} 

\subsubsection{Spherical with direction cosines}

In $d$-dimensions there are $n$ orthogonal $2$-planes~\footnote{Note that this is linked to the fact that the little group of massive representation in $D$ dimension is $\group{SO}(N)$, which possess $n$ Casimir invariants~\cite{myers_black_1986}.}, thus we can pair $2n$ of the embedding coordinates $X^a$ \eqref{coord:eq:spherical-embedding} as $(X_i, Y_i)$ which are parametrized as
\begin{equation}
	X_i + i Y_i = \mu_i \e^{i\phi_i}, \qquad
	a = 1, \ldots n.
\end{equation} 
For $d$ even there is an extra unpaired coordinate that is taken to be
\begin{equation}
	X^N = \alpha.
\end{equation}

Each pair parametrizes a $2$-sphere of radius $\mu_i$.
The $\mu_i$ are called the \emph{direction cosines} and satisfy
\begin{equation}
	\sum_i \mu_i^2 + \varepsilon' \alpha^2 = 1
\end{equation} 
since there is one superfluous coordinate from the embedding.

Finally the metric is
\begin{equation}
	\dd \Omega_{N-1}^2 = \sum_i \Big(\dd \mu_i^2 + \mu_i^2\; \dd \phi_i^2 \Big) + \varepsilon'\, \dd \alpha^2.
\end{equation} 

The interest of these coordinates is that all rotational directions are symmetric.

\subsubsection{Spheroidal with direction cosines}
\label{app:coord:general-d:oblate-cosines}

From the previous system we can define the spheroidal $(\bar r, \bar\mu_i, \bar\phi_i)$ system – adapted when some of the $2$-spheres are deformed to ellipses – by introducing parameters $a_i$ such that (for $d$ odd)
\begin{equation}
	\label{coord:eq:spherical-to-oblate-mu}
	r^2 \mu_i^2 = (\bar r^2 + a_i^2) \bar \mu_i^2, \qquad
	\sum_i \bar \mu_i^2 = 1.
\end{equation} 
This last condition implies that
\begin{equation}
	r^2 = \sum_i (\bar r^2 + a_i^2) \bar \mu_i^2
		= \bar r^2 + \sum_i a_i^2 \bar \mu_i^2.
\end{equation} 

In these coordinates the metric reads
\begin{equation}
	\label{coord:metric:flat-d:spheroidal}
	\dd \Sigma^2 = F\; \dd \bar r^2 + \sum_i (\bar r^2 + a_i^2) \Big(\dd \bar \mu_i^2 + \bar \mu_i^2\; \dd \bar \phi_i^2 \Big) + \varepsilon'\, r^2 \dd \alpha^2
\end{equation} 
and we defined
\begin{equation}
	\label{coord:eq:flat-d:spheroidal:F}
	F = 1 - \sum_i \frac{a_i^2 \bar \mu_i^2}{\bar r^2 + a_i^2} = \sum_i \frac{\bar r^2 \bar \mu_i^2}{\bar r^2 + a_i^2}.
\end{equation} 

Here the $a_i$ are just introduced as parameters in the transformation, but in the main text they are interpreted as "true" rotation parameters, i.e.
angular momenta (per unit of mass) of a black hole.
They all appear on an equal footing.

Another quantity of interest is
\begin{equation}
	\label{coord:eq:flat-d:spheroidal:Pi}
	\Pi = \prod_i (\bar r^2 + a_i^2).
\end{equation} 

\subsubsection{Mixed spherical–spheroidal}
\label{app:coord:general-d:oblate-spherical}

We consider the deformation of the spherical metric where one of the $2$-sphere is replaced by an ellipse~\cite[sec.~3]{aliev_rotating_2006}.

To shorten the notation let's define
\begin{equation}
	\theta = \theta_{N-1}, \qquad
	\phi = \theta_{N-2}.
\end{equation} 
Doing the change of coordinates
\begin{equation}
	\sin^2 \theta \sin^2 \phi = \cos^2 \theta.
\end{equation}
the metric changes to
\begin{equation}
	\dd \Sigma^2 = \frac{\rho^2}{r^2 + a^2}\, \dd r^2
		+ \rho^2 \dd\theta^2 \\
		+ (r^2 + a^2)\, \sin^2 \theta\, \dd\phi^2
		+ r^2 \cos^2 \theta^2\, \dd\Omega_{d-4}^2
\end{equation} 
where as usual
\begin{equation}
	\rho^2 = r^2 + a^2 \cos^2 \theta.
\end{equation} 
Except for the last term one recognize $4$-dimensional oblate spheroidal coordinates \eqref{coord:metric:4d:spheroidal}.

\subsection{4-dimensional}
\label{app:coord:4d}

In this section one consider
\begin{equation}
	d = 4, \quad
	N = 3, \quad
	n = 1.
\end{equation} 

\subsubsection{Cartesian system}

\begin{equation}
	\dd \Sigma^2 = \dd x^2 + \dd y^2 + \dd z^2
\end{equation} 

\subsubsection{Spherical}

\begin{subequations}
\begin{gather}
	\dd \Sigma^2 = \dd r^2 + r^2 \dd \Omega^2, \\
	\dd \Omega^2 = \dd \theta^2 + \sin^2 \theta\; \dd \phi^2,
\end{gather}
\end{subequations}
where $\dd \Omega^2 \equiv \dd \Omega_2^2$.

\subsubsection{Spherical with direction cosines}

\begin{subequations}
\begin{gather}
	\dd \Omega^2 = \dd \mu^2 + \mu^2\; \dd \phi^2 + \dd \alpha^2, \\
	\mu^2 + \alpha^2 = 1,
\end{gather}
\end{subequations}
where
\begin{equation}
	x + iy = r \mu\, \e^{i\phi}, \qquad
	z = r \alpha,
\end{equation} 

Using the constraint one can rewrite
\begin{equation}
	\dd \Omega^2 = \frac{1}{1 - \mu^2}\; \dd \mu^2 + \mu^2\; \dd \phi^2.
\end{equation} 
Finally the change of coordinates
\begin{equation}
	\alpha = \cos \theta, \qquad
	\mu = \sin \theta.
\end{equation} 
solves the constraint and gives back the spherical coordinates.

\subsubsection{Spheroidal with direction cosines}

The oblate spheroidal coordinates from the Cartesian ones are~\cite[p.~15]{visser_kerr_2007}
\begin{equation}
	x + iy = \sqrt{r^2 + a^2}\, \sin \theta\, \e^{i\phi}, \qquad
	z = r \cos\theta,
\end{equation} 
and the metric is
\begin{equation}
	\label{coord:metric:4d:spheroidal}
	\dd \Sigma^2 = \frac{\rho^2}{r^2 + a^2}\; \dd r^2 + \rho^2 \dd\theta^2 + (r^2 + a^2) \sin^2 \theta\; \dd \phi^2, \qquad
	\rho^2 = r^2 + a^2 \cos^2 \theta.
\end{equation} 

In terms of direction cosines one has
\begin{equation}
	\dd \Sigma^2 = \left(1 - \frac{r^2 \mu^2}{r^2 + a^2} \right)\; \dd r^2 + (r^2 + a^2) \Big(\dd \mu^2 + \mu^2\; \dd \phi^2 \Big) + r^2 \dd \alpha^2.
\end{equation} 

\subsection{5-dimensional}
\label{app:coord:5d}

In this section one consider
\begin{equation}
	d = 4, \quad
	N = 3, \quad
	n = 1.
\end{equation} 

\subsubsection{Spherical with direction cosines}

\begin{equation}
	\label{coord:metric:5d:spherical}
	\dd\Omega_3^2 = \dd \mu^2 + \mu^2\, \dd\phi^2 + \dd \nu^2 + \nu^2\, \dd\psi^2, \qquad
	\mu^2 + \nu^2 = 1
\end{equation} 
where for simplicity
\begin{equation}
	\mu = \mu_1, \qquad
	\mu = \mu_2, \qquad
	\phi = \phi_1, \qquad
	\psi = \phi_2.
\end{equation}

\subsubsection{Hopf coordinates}
\label{app:coord:5d:hopf}

The constraint \eqref{coord:metric:5d:spherical} can be solved by
\begin{equation}
	\mu = \sin \theta, \qquad
	\nu = \cos \theta
\end{equation} 
and this gives the metric in Hopf coordinates
\begin{equation}
	\label{coord:metric:5d:hopf}
	\dd \Omega_3^2 = \dd\theta^2 + \sin^2 \theta\, \dd\phi^2 + \cos^2 \theta\, \dd\psi^2.
\end{equation} 

\section{Janis–Newman algorithm for any dimension}
\label{app:higher-dim-jna}

In this appendix we consider the JN algorithm applied to a general static $d$-dimension metric.
As we argued in a previous section it is important to consider separately the transformation of the metric and the complexification of the functions inside.
Hence we are able to derive the general form of a rotating metric with the maximal number of angular momenta it can have in $d$ dimensions, but we are not able to apply this result to any specific example for $d \ge 6$, except if all momenta but one are vanishing~\cite{xu_exact_1988}.
Despite this last problem, this computation provides a unified framework for $d = 3, 4, 5$ (see appendix~\ref{sec:BTZ} for the BTZ black hole).

In the following the dimension is taken to be odd in order to simplify the computations, but the final result holds also for $d$ even.

\subsection{Metric transformation}

\subsubsection{Seed metric and discussion}

Consider the $d$-dimensional static metric (notations are defined in appendix~\ref{app:coord:general-d})
\begin{equation}
	\dd s^2 = - f\, \dd t^2 + f^{-1}\, \dd r^2 + r^2\, \dd \Omega_{d-2}^2
\end{equation} 
where $\dd \Omega_{d-2}^2$ is the metric on $S^{d-2}$
\begin{equation}
	\dd \Omega_{d-2}^2 = \dd\theta_{d-2} + \sin^2 \theta_{d-2}\, \dd \Omega_{d-3}^2
		= \sum_{i=1}^n \big( \dd\mu_i^2 + \mu_i^2 \dd\phi_i^2).
\end{equation} 
The number $n = (d-1) / 2$ denotes the number of independent $2$-spheres.

In Eddington–Finkelstein coordinates the metric reads
\begin{equation}
	\label{higher-jna:metric:static-seed}
	\dd s^2 = (1 - f)\, \dd u^2 - \dd u\, (\dd u + 2 \dd r)
			+ r^2 \sum_i \Big(\dd \mu_i^2 + \mu_i^2\; \dd \phi_i^2 \Big).
\end{equation} 

The metric looks like a $2$-dimensional space $(t, r)$ with a certain number of additional $2$-spheres $(\mu_i, \phi_i)$ which are independent from one another.
Then we can consider only the piece $(u, r, \mu_i, \phi_i)$ (for fixed $i$) which will transform like a $4$-dimensional spacetime, while the other part of the metric $(\mu_j, \phi_j)$ for all $j \neq i$ will be unchanged.
After the first transformation we can move to another $2$-sphere.
We can thus imagine to put in rotation only one of these spheres.
Then we will apply again and again the algorithm until all the spheres have angular momentum: the whole complexification will thus be a $n$-steps process.
Moreover if these $2$-spheres are taken to be independent this implies that we should not complexify the functions that are not associated with the plane we are putting in rotation.

To match these demands the metric is rewritten as
\begin{equation}
	\label{higher-jna:metric:static-seed-ur}
	\dd s^2 = (1 - f)\, \dd u^2 - \dd u\, (\dd u + 2 \dd r_{i_1})
		+ r_{i_1}^2 (\dd\mu_{i_1}^2 + \mu_{i_1}^2 \dd\phi_{i_1}^2)
		+ \sum_{i \neq i_1} \Big(r_{i_1}^2 \dd \mu_i^2 + R^2 \mu_i^2\; \dd \phi_i^2 \Big).
\end{equation} 
where we introduced the following two functions of $r$
\begin{equation}
	r_{i_1}(r) = r, \qquad R(r) = r\,.
\end{equation} 
This allows to choose different complexification for terms in front of different terms in the metric.
It may be surprising to note that the factors in front of $\dd \mu_i^2$ have been chosen to be $r_{i_1}^2$ and not $R^2$, but the reason is that the $\mu_i$ are all linked by the constraint
\begin{equation}
	\sum_i \mu_i^2 = 1
\end{equation} 
and the transformation of one $i_1$-th $2$-sphere will change the corresponding $\mu_{i_1}$, but also all the others, as it is clear from the formula \eqref{coord:eq:spherical-to-oblate-mu} with all the $a_i$ vanishing but one (this can also be observed in $5d$ where both $\mu_i$ are gathered into $\theta$).

\subsubsection{First transformation}

The transformation is chosen to be
\begin{subequations}
\label{higher-jna:change:jna-1}
\begin{equation}
	r_{i_1} = r'_{i_1} - i\, a_{i_1} \sqrt{1 - \mu_{i_1}^2}, \qquad
	u = u' + i\, a_{i_1} \sqrt{1 - \mu_{i_1}^2}
\end{equation} 
which, together with the ansatz
\begin{equation}
	i\; \frac{\dd \mu_{i_1}}{\sqrt{1 - \mu_{i_1}^2}} = \mu_{i_1}\, \dd \phi_{i_1},
\end{equation} 
gives the differentials
\begin{equation}
	\dd r_{i_1} = \dd r'_{i_1} + a_{i_1} \mu_{i_1}^2\, \dd \phi_{i_1}, \qquad
	\dd u = \dd u' - a_{i_1} \mu_{i_1}^2\, \dd \phi_{i_1}.
\end{equation} 
\end{subequations}

It is easy to check that this transformation reproduces the one given in four and five dimensions.

The complexified version of $f$ is written as $\tilde f^{\{i_1\}}$: we need to keep track of the order in which we gave angular momentum since the function $\tilde f$ will be transformed at each step.

We consider separately the transformation of the $(u, r)$ and $\{ \mu_i, \phi_i \}$ parts.
Inserting the transformations \eqref{higher-jna:change:jna-1} in \eqref{higher-jna:metric:static-seed} results in
\begin{subequations}
\begin{align*}
	\dd s_{u,r}^2 &= (1 - \tilde f^{\{i_1\}})\, \Big(\dd u - a_{i_1} \mu_{i_1}^2\, \dd \phi_{i_1} \Big)^2
		- \dd u\, (\dd u + 2 \dd r_{i_1})
		+ 2 a_{i_1} \mu_{i_1}^2\, \dd r_{i_1} \dd \phi_{i_1}
		+ a_{i_1}^2 \mu_{i_1}^4\, \dd \phi_{i_1}^2, \\
	\dd s_{\mu,\phi}^2 &= \big( r_{i_1}^2 + a_{i_1}^2 \big) (\dd\mu_{i_1}^2 + \mu_{i_1}^2 \dd\phi_{i_1}^2)
		+ \sum_{i \neq i_1} \big( r_{i_1}^2 \dd \mu_i^2 + R^2 \mu_i^2\, \dd \phi_i^2 \big) - a_{i_1}^2 \mu_{i_1}^4\, \dd \phi_{i_1}^2 \\
		&\qquad + a_{i_1}^2 \bigg[- \mu_{i_1}^2 \dd \mu_{i_1}^2 + (1 - \mu_{i_1}^2) \sum_{i \neq i_1} \dd \mu_i^2 \bigg].
\end{align*}
\end{subequations}

The term in the last bracket vanishes as can be seen by using the differential of the constraint
\begin{equation}
	\sum_i \mu_i^2 = 1 \Longrightarrow
	\sum_i \mu_i \dd\mu_i = 0.
\end{equation} 
Since this step is very important and non-trivial we expose the details
\begin{align*}
	[\cdots] &= \mu_{i_1}^2 \dd \mu_{i_1}^2 - (1 - \mu_{i_1}^2) \sum_{i \neq i_1} \dd \mu_i^2
		= \left(\sum_{i \neq i_1} \mu_i \dd\mu_i \right)^2 - \sum_{j \neq i_1} \mu_j^2 \sum_{i \neq i_1} \dd \mu_i^2 \\
		&= \sum_{i,j \neq i_1} \big(\mu_i \mu_j \dd\mu_i \dd\mu_j - \mu_j^2 \dd \mu_i^2 \big)
		= \sum_{i,j \neq i_1} \mu_j \big(\mu_i \dd\mu_j - \mu_j \dd \mu_i \big) \dd\mu_i
		= 0
\end{align*}
by antisymmetry.

Setting $r_{i_1} = R = r$ one obtains the metric
\begin{equation}
\begin{aligned}
	\dd s^2 &= (1 - \tilde f^{\{i_1\}})\, \Big(\dd u - a_{i_1} \mu_{i_1}^2\, \dd \phi_{i_1} \Big)^2
		- \dd u\, (\dd u + 2 \dd r)
		+ 2 a_{i_1} \mu_{i_1}^2\, \dd r \dd \phi_{i_1} \\
		&\qquad+ \big( r^2 + a_{i_1}^2 \big) (\dd\mu_{i_1}^2 + \mu_{i_1}^2 \dd\phi_{i_1}^2)
		+ \sum_{i \neq i_1} r^2 \big( \dd \mu_i^2 + \mu_i^2\, \dd \phi_i^2 \big).
\end{aligned}
\end{equation}
It corresponds to Myers–Perry metric in $d$ dimensions with one non-vanishing angular momentum.
We recover the same structure as in \eqref{higher-jna:metric:static-seed-ur} with some extra terms that are specific to the $i_1$-th $2$-sphere.

\subsubsection{Iteration and final result}

We should now split again $r$ in functions $(r_{i_2}, R)$.
Very similarly to the first time we have
\begin{equation}
\begin{aligned}
	\dd s^2 &= (1 - \tilde f^{\{i_1\}})\, \Big(\dd u - a_{i_1} \mu_{i_1}^2\; \dd \phi_{i_1} \Big)^2
		- \dd u\, (\dd u + 2 \dd r_{i_2})
		+ 2 a_{i_1} \mu_{i_1}^2\; \dd R \dd \phi_{i_1} \\
		&\qquad+ \big( r_{i_2}^2 + a_{i_1}^2 \big) \dd\mu_{i_1}^2
		+ \big( R^2 + a_{i_1}^2 \big) \mu_{i_1}^2 \dd\phi_{i_1}^2
		+ r_{i_2}^2 ( \dd\mu_{i_2}^2 + \mu_{i_2}^2 \dd\phi_{i_2}^2 ) \\
		&\qquad+ \sum_{i \neq i_1, i_2} \Big(r_{i_2}^2 \dd \mu_i^2 + R^2 \mu_i^2\; \dd \phi_i^2 \Big).
\end{aligned}
\end{equation}

We can now complexify as
\begin{equation}
	r_{i_2} = r'_{i_2} - i a_{i_2} \sqrt{1 - \mu_{i_2}^2}, \qquad
	u = u' + i\, a_{i_1} \sqrt{1 - \mu_{i_2}^2}.
\end{equation} 
The steps are exactly the same as before, except that we have some inert terms.
The complexified functions is now $\tilde f^{\{i_1, i_2\}}$.

Repeating the procedure $n$ times we arrive at
\begin{equation}
	\label{metric:rotating-result-jna-ur}
	\begin{aligned}
		\dd s^2 = &- \dd u^2 - 2 \dd u \dd r
			+ \sum_i (r^2 + a_i^2) (\dd \mu_i^2 + \mu_i^2 \dd \phi_i^2)
			- 2 \sum_i a_i \mu_i^2 \; \dd r \dd \phi_i \\
			&+ \Big(1 - \tilde f^{\{i_1, \ldots, i_n\}} \Big) \left(\dd u + \sum_i a_i \mu_i^2 \dd \phi_i \right)^2.
	\end{aligned}
\end{equation} 
One recognizes the general form of the $d$-dimensional metric with $n$ angular momenta~\cite{myers_black_1986}.

Let's quote the metric in Boyer–Lindquist coordinates (omitting the indices on $\tilde f$)~\cite{myers_black_1986}
\begin{equation}
	\label{metric:rotating-result-jna-bl}
	\dd s^2 = - \dd t^2
		+ (1 - \tilde f) \left(\dd t - \sum_i a_i \mu_i^2 \dd \phi_i \right)^2
		+ \frac{r^2 \rho^2}{\Delta}\; \dd r^2
		+ \sum_i (r^2 + a_i^2) \Big(\dd \mu_i^2 + \mu_i^2\; \dd \phi_i^2 \Big)
\end{equation} 
which is obtained from the transformation
\begin{subequations}
\label{change:rotating:higher-dim-func-gh}
\begin{align}
	g &= \frac{\Pi}{\Delta}
		= \frac{1}{1 - F (1 - \tilde f)}, \\
	h_i &= \frac{\Pi}{\Delta} \; \frac{a_i}{r^2 + a_i^2}.
\end{align}
\end{subequations}
where the various quantities involved are (see appendix~\ref{app:coord:general-d:oblate-cosines})
\begin{equation}
	\label{metric:rotating-result-jna-bl-parameters}
	\begin{gathered}
		\Pi = \prod_i (r^2 + a_i^2), \qquad
		F = 1 - \sum_i \frac{a_i^2 \mu_i^2}{r^2 + a_i^2} = r^2 \sum_i \frac{\mu_i^2}{r^2 + a_i^2}, \\
		r^2 \rho^2 = \Pi F, \qquad
		\Delta = \tilde f\, r^2 \rho^2 + \Pi (1 - F).
	\end{gathered}
\end{equation}

Before ending this section, we comment the case of even dimensions: the term $\varepsilon'\, r^2 \dd \alpha^2$ is complexified as $\varepsilon'\, r_{i_1}^2 \dd \alpha^2$, since it contributes to the sum
\begin{equation}
	\sum_i \mu_i^2 + \alpha^2 = 1.
\end{equation} 
This can be seen more clearly by defining $\mu_{n+1} = \alpha$ (we can also define $\phi_{n+1} = 0$), in which case the index $i$ runs from $1$ to $n+\varepsilon$, and all the previous computations are still valid.

\subsection{Examples}

\subsubsection{Application to flat space}

A first and trivial example is to take $f = 1$.
In this case one recovers Minkowski metric in spheroidal coordinates with direction cosines (appendix~\ref{app:coord:general-d:oblate-cosines})
\begin{equation}
	\dd s^2 = - \dd t^2 + F\; \dd \bar r^2 + \sum_i (\bar r^2 + a_i^2) \Big(\dd \bar \mu_i^2 + \bar \mu_i^2\; \dd \bar \phi_i^2 \Big) + \varepsilon'\, r^2 \dd \alpha^2.
\end{equation}
In this case the JN algorithm is equivalent to a (true) change of coordinates~\cite{ferraro_untangling_2014} and there is no intrinsic rotation.
The presence of a non-trivial function $f$ then deforms the algorithm.

\subsubsection{Myers–Perry with one angular momentum}

The derivation of the Myers–Perry metric with one non-vanishing angular momentum has been found by Xu~\cite{xu_exact_1988}.

The transformation is taken to be in the first plane
\begin{equation}
	r = r' - i a \sqrt{1 - \mu^2}
\end{equation} 
where $\mu \equiv \mu_1$.
The transformation to the mixed spherical–spheroidal system (appendix~\ref{app:coord:general-d:oblate-spherical} is obtained by setting
\begin{equation}
	\mu = \sin \theta, \qquad
	\phi_1 = \phi.
\end{equation} 
In these coordinates the transformation reads
\begin{equation}
	r = r' - i a \cos \theta.
\end{equation} 
We will use the quantity
\begin{equation}
	\rho^2 = r^2 + a^2 (1 - \mu^2)
		= r^2 + a^2 \cos^2 \theta.
\end{equation} 

The Schwarzschild–Tangherlini metric is~\cite{tangherlini_schwarzschild_1963}
\begin{equation}
	\dd s^2 = - f\, \dd t^2 + f^{-1}\, \dd r^2 + r^2\, \dd \Omega_{d-2}^2, \qquad
	f = 1 - \frac{m}{r^{d-3}}.
\end{equation} 

Applying the previous transformation results in
\begin{equation}
\begin{aligned}
	\dd s^2 &= (1 - \tilde f)\, \Big(\dd u - a \mu^2\, \dd \phi \Big)^2
		- \dd u\, (\dd u + 2 \dd r)
		+ 2 a \mu^2\, \dd r \dd \phi \\
		&\qquad+ \big( r^2 + a^2 \big) (\dd\mu^2 + \mu^2 \dd\phi^2)
		+ \sum_{i \neq 1} r^2 \big( \dd \mu_i^2 + \mu_i^2\, \dd \phi_i^2 \big).
\end{aligned}
\end{equation}
where $f$ has been complexified as
\begin{equation}
	\tilde f = 1 - \frac{m}{\rho^2 r^{d-5}}.
\end{equation} 

In the mixed coordinate system one has~\cite{xu_exact_1988, aliev_rotating_2006}
\begin{equation}
	\begin{aligned}
		\dd s^2 = &- \tilde f\, \dd t^2
			+ 2 a (1 - \tilde f) \sin^2 \theta\, \dd t \dd\phi
			+ \frac{r^{d-3} \rho^2}{\Delta}\, \dd r^2 + \rho^2 \dd\theta^2 \\
			&+ \frac{\Sigma^2}{\rho^2}\, \sin^2 \theta\, \dd\phi^2
			+ r^2 \cos^2 \theta^2\, \dd\Omega_{d-4}^2.
	\end{aligned}
\end{equation} 
where we defined as usual
\begin{equation}
	\Delta = \tilde f \rho^2 + a^2 \sin^2 \theta, \qquad
	\frac{\Sigma^2}{\rho^2} = r^2 + a^2 + a g_{t\phi}.
\end{equation} 

This last expression explains why the transformation is straightforward with one angular momentum: the transformation is exactly the one for $d = 4$ and the extraneous dimensions are just spectators.

We have not been able to generalize this result for several non-vanishing momenta for $d \ge 6$, even for the case with equal momenta.

\subsubsection{BTZ black hole}
\label{sec:BTZ}

As another application we show how to derive the $d=3$ rotating BTZ black hole from its static version~\cite{banados_black_1992}
\begin{equation}
	\dd s^2 = - f\, \dd t^2 + f^{-1}\, \dd r^2 + r^2 \dd\phi^2, \qquad
	f(r) = - M + \frac{r^2}{\ell^2},
\end{equation} 
where $\Lambda = - 1 / \ell^2$ is the cosmological constant.

In three dimensions the metric on $S^1$ in spherical coordinates is given by
\begin{equation}
	\dd\Omega_1^2 = \dd\phi^2.
\end{equation} 
Introducing the coordinate $\mu$ we can write it in oblate spheroidal coordinates
\begin{equation}
	\dd\Omega_1^2 = \dd\mu^2 + \mu^2 \dd\phi^2
\end{equation} 
with the constraint
\begin{equation}
	\mu^2 = 1.
\end{equation} 

Application of the transformation
\begin{equation}
	u = u' + i a \sqrt{1 - \mu^2}, \qquad
	r = r' - i a \sqrt{1 - \mu^2}
\end{equation} 
gives from \eqref{metric:rotating-result-jna-ur}
\begin{equation}
	\begin{aligned}
		\dd s^2 = &- \dd u^2 - 2 \dd u \dd r
			+ (r^2 + a^2) (\dd \mu^2 + \mu^2 \dd \phi^2)
			- 2 a \mu^2 \; \dd r \dd \phi \\
			&+ (1 - \tilde f) (\dd u + a \mu^2 \dd \phi )^2.
	\end{aligned}
\end{equation} 
We still need to give the complexification of $f$ which is
\begin{equation}
	\tilde f = - M + \frac{\rho^2}{\ell^2}, \qquad
	\rho^2 = r^2 + a^2 (1 - \mu^2).
\end{equation} 

The transformation \eqref{change:rotating:higher-dim-func-gh}
\begin{equation}
	g = \frac{\rho^2 (1 - \tilde f)}{\Delta}, \qquad
	h = \frac{a}{\Delta}, \qquad
	\Delta = r^2 + a^2 + (\tilde f - 1) \rho^2
\end{equation}
to Boyer–Lindquist coordinates leads to the metric \eqref{metric:rotating-result-jna-bl}
\begin{equation}
	\dd s^2 = - \dd t^2
		+ (1 - \tilde f) (\dd t + a \mu^2 \dd \phi )^2
		+ \frac{\rho^2}{\Delta}\; \dd r^2
		+ (r^2 + a^2) (\dd \mu^2 + \mu^2\; \dd \phi^2 ).
\end{equation} 

Finally we can use the constraint $\mu^2 = 1$ to remove the $\mu$.
In this case we have
\begin{equation}
	\rho^2 = r^2, \qquad
	\Delta = a^2 + \tilde f r^2
\end{equation}
and the metric simplifies to
\begin{equation}
	\dd s^2 = - \dd t^2
		+ (1 - \tilde f) (\dd t + a \dd \phi )^2
		+ \frac{r^2}{a^2 + r^2 \tilde f}\; \dd r^2
		+ (r^2 + a^2) \dd \phi^2.
\end{equation} 

We define the function
\begin{equation}
	N^2 = \tilde f + \frac{a^2}{r^2} = - M + \frac{r^2}{\ell^2} + \frac{a^2}{r^2}.
\end{equation} 
Then redefining the time variable as~\cite{kim_notes_1997, kim_spinning_1999}
\begin{equation}
	t = t' - a \phi
\end{equation} 
we get (omitting the prime)
\begin{equation}
	\dd s^2 = - N^2 \dd t^2 + N^{-2}\, \dd r^2 + r^2 (N^\phi \dd t + \dd \phi)^2
\end{equation} 
with the angular shift
\begin{equation}
	N^\phi(r) = \frac{a}{r^2}.
\end{equation} 
This is the solution given in~\cite{banados_black_1992} with $J = -2a$.

This has already been done by Kim~\cite{kim_notes_1997, kim_spinning_1999} in a different settings: he views the $d=3$ solution as the slice $\theta = \pi/2$ of the $d=4$ solution.
Obviously this is equivalent to our approach: we have seen that $\mu = \sin \theta$ in $d=4$ (appendix~\ref{app:coord:4d}), and the constraint $\mu^2 = 1$ is solved by $\theta = \pi/2$.
Nonetheless our approach is more direct since the result just follows from a suitable choice of coordinates and there is no need for advanced justification.

Starting from the charged BTZ black hole
\begin{equation}
	f(r) = - M + \frac{r^2}{\ell^2} - Q^2 \ln r^2, \qquad
	A = - \frac{Q}{2}\, \ln r^2,
\end{equation} 
it is not possible to find the charged rotating BTZ black hole from~\cite[sec.~4.2]{martinez_charged_2000}: the solution solves Einstein equations, but not the Maxwell ones.
It may be possible that a more general ansatz is necessary (see~\cite{erbin_deciphering_2014}).
This has been already remarked using another technique in~\cite[app.~B]{lambert_conformal_2014}.

\subsubsection{Five-dimensional Myers–Perry}
\label{sec:higher-dim:examples:MP-5d}

We take a new look at the five-dimensional Myers–Perry solution in order to derive it in spheroidal coordinates because it is instructive.

The function
\begin{equation}
	1 - f = \frac{m}{r^2}
\end{equation} 
is first complexified as
\begin{equation}
	1 - \tilde f^{\{1\}} = \frac{m}{\abs{r_1}^2}
		= \frac{m}{r^2 + a^2 (1 - \mu^2)}
\end{equation}
and then as 
\begin{equation}
	1 - \tilde f^{\{1, 2\}} = \frac{m}{\abs{r_2}^2 + a^2 (1 - \mu^2)}
		= \frac{m}{r^2 + a^2 (1 - \mu^2) + b^2 (1 - \nu^2)}.
\end{equation}
after the two transformations
\begin{equation}
	r_1 = r_1' - i a \sqrt{1 - \mu^2}, \qquad
	r_2 = r_2' - i b \sqrt{1 - \nu^2}.
\end{equation} 
For $\mu = \sin \theta$ and $\nu = \cos \theta$ one recovers the transformations from sections~\ref{sec:MP-5d} and \ref{sec:BMPV}.

Let's denote the denominator by $\rho^2$ and compute
\begin{align*}
	\frac{\rho^2}{r^2} &= r^2 + a^2 (1 - \mu^2) + b^2 (1 - \nu^2)
		= (\mu^2 + \nu^2) r^2 + \nu^2 a^2 + \mu^2 b^2 \\
		&= \mu^2 (r^2 + b^2) + \nu^2 (r^2 + a^2)
		= (r^2 + b^2) (r^2 + a^2) \left( \frac{\mu^2}{r^2 + a^2} + \frac{\nu^2}{r^2 + b^2} \right).
\end{align*}
and thus
\begin{equation}
	r^2 \rho^2 = \Pi F.
\end{equation} 
Plugging this into $\tilde f^{\{1, 2\}}$ we have~\cite{myers_black_1986}
\begin{equation}
	1 - \tilde f^{\{1, 2\}} = \frac{m r^2}{\Pi F}.
\end{equation} 

\section{Another approach to BMPV}
\label{app:bmpv-second-approach}

In section~\ref{sec:BMPV} we applied the same recipe given in section~\ref{sec:MP-5d} which, according to our claim, is the standard procedure in five dimensions.

There is another way to derive BMPV black hole.
Indeed, by considering that terms quadratic in the angular momentum will not survive the extremal limit, they can be added the metric without modifying the final result.
Hence we can decide to transform all the terms of the metric~\footnotemark{} since the additional terms will be subleading.
As a result the BL transformation is directly well defined and overall formulas are simpler, but we need to take the extremal limit before the end (this could be done either in $(u, r)$ or $(t, r)$ coordinates).
This appendix shows that both approaches give the same result.
\footnotetext{In opposition to our initial recipe, but this is done in an controlled way.}

Applying the two transformations
\begin{subequations}
\begin{gather}
	u = u' + i a \cos \theta, \qquad
	\dd u = \dd u' - a \sin^2 \theta\, \dd\phi, \\
	u = u' + i a\, \sin \theta, \qquad
	\dd u = \dd u' - a \cos^2 \theta\, \dd\psi
\end{gather}
\end{subequations}
successively on all the terms one obtains the metric
\begin{equation}
	\begin{aligned}
		\dd s^2 = &- \tilde H^{-2} \big(\dd u
				- a (1 - \tilde H^{3/2}) (\sin^2 \theta\, \dd\phi + \cos^2 \theta\, \dd\psi) \big)^2 \\
			&- 2 \tilde H^{-1/2} \big(\dd u - a (\sin^2 \theta\, \dd\phi + \cos^2 \theta\, \dd\psi) \big)\, \dd r \\
			&+ \tilde H\, \Big(
				(r^2 + a^2) (\dd \theta^2 + \sin^2 \theta\, \dd\phi^2 + \cos^2 \theta\, \dd\psi^2)
				+ a^2 (\sin^2 \theta\, \dd\phi + \cos^2 \theta\, \dd\psi)^2 \Big),
	\end{aligned}
\end{equation} 
where again $\tilde H$ is given by \eqref{eq:5d-bmpv:tilde-H}
\begin{equation}
	\tilde H = 1 + \frac{m}{r^2 + a^2}.
\end{equation} 
Only one term is different when comparing with \eqref{metric:5d-bmpv:bmpv-metric-ur-before-limit}.

The BL transformation \eqref{eq:5d-bl-transformation} is well-defined and the corresponding functions are
\begin{equation}
	\label{eq:5d-bl-transformation}
	g(r) = \frac{a^2 + (r^2 + a^2) \tilde H(r)}{r^2 + 2 a^2}, \qquad
	h_\phi(r) = h_\psi(r) = \frac{a}{r^2 + 2 a^2}
\end{equation} 
which do not depend on $\theta$.
They lead to the metric
\begin{equation}
	\begin{aligned}
		\dd s^2 = &- \tilde H^{-2} \big(\dd t
				- a (1 - \tilde H^{3/2}) (\sin^2 \theta\, \dd\phi + \cos^2 \theta\, \dd\psi) \big)^2 \\
			&+ \tilde H\, \bigg[
				(r^2 + a^2) \left(\frac{\dd r^2}{r^2 + 2 a^2} + \dd \theta^2 + \sin^2 \theta\, \dd\phi^2 + \cos^2 \theta\, \dd\psi^2 \right) \\
				&\qquad\quad+ a^2 (\sin^2 \theta\, \dd\phi + \cos^2 \theta\, \dd\psi)^2 \bigg].
	\end{aligned}
\end{equation} 

At this point it is straightforward to check that this solution does not satisfy Einstein equations and we need to take the extremal limit \eqref{eq:5d-bmpv-extremal-limit}
\begin{equation}
	a, m \longrightarrow 0, \qquad
	\text{imposing} \qquad
	\frac{m}{a^2} = \cst
\end{equation}
in order to get the BMPV solution \eqref{metric:5d-bmpv:bmpv-metric}
\begin{equation}
	\begin{aligned}
		\dd s^2 = &- \tilde H^{-2} \left(\dd t
				+ \frac{3\, m a}{2\, r^2}\, (\sin^2 \theta\, \dd\phi + \cos^2 \theta\, \dd\psi) \right)^2 \\
			&+ \tilde H\, \Big(\dd r^2 + r^2 \big( \dd \theta^2 + \sin^2 \theta\, \dd\phi^2 + \cos^2 \theta\, \dd\psi^2 \big) \Big).
	\end{aligned}
\end{equation} 

It is surprising that the BL transformation is simpler in this case.
Another point that is worth of stressing is that we did not need to take the extremal limit in this computation, whereas in section~\ref{sec:BMPV} we had to in order to get a well-defined BL transformation.

\printbibliography[heading=bibintoc]

\end{document}